\documentclass[final,5p,times,twocolumn]{elsarticle}

\usepackage{lineno,hyperref}
\usepackage[greek,english]{babel}
\modulolinenumbers[5]

\journal{Acta Astronautica}

\newcommand{\new}{}

\begin{document}

\begin{frontmatter}

\title{Concepts for future missions to search for technosignatures}

\author{Hector Socas-Navarro}
\address{Instituto de Astrof\' \i sica de Canarias, C/V\' \i a L\'actea s/n, E-38205 La Laguna, Tenerife, Spain\\
Departamento de Astrof\' \i sica, Universidad de La Laguna, E-38206 La Laguna, Tenerife, Spain}

\author{Jacob Haqq-Misra}
\address{Blue Marble Space Institute of Science, Seattle, WA, USA}

\author{Jason T.\ Wright}
\address{Department of Astronomy \& Astrophysics, 525 Davey Laboratory, The Pennsylvania State University, University Park, PA, 16802, USA\\
Center for Exoplanets and Habitable Worlds, 525 Davey Laboratory, The Pennsylvania State University, University Park, PA, 16802, USA\\
Penn State Extraterrestial Intelligence Center, 525 Davey Laboratory, The Pennsylvania State University, University Park, PA, 16802, USA}

\author{Ravi Kopparapu}
\address{NASA Goddard Space Flight Center, 8800 Greenbelt Road, Greenbelt, MD 20771, USA}

\author{James Benford}
\address{Microwave Sciences, Lafayette, CA, USA}

\author{Ross Davis}
\address{Indiana University, Indianapolis, IN, USA}

\author{and the TechnoClimes 2020 workshop participants}




\begin{abstract}
New and unique opportunities now exist to look for technosignatures (TS) beyond traditional SETI radio searches, motivated by tremendous advances in exoplanet science and observing capabilities in recent years. Space agencies, both public and private, may be particularly interested in learning about the community's views as to the optimal methods for future TS searches with current or forthcoming technology. This report is an effort in that direction. We put forward a set of possible mission concepts designed to search for TS, although the data supplied by such missions would also benefit other areas of astrophysics. We introduce a novel framework to \new{analyze} a broad diversity of TS in a quantitative manner. \new{This framework is based on} the concept of {\em ichnoscale}, \new{which is a new parameter related to the scale of a TS cosmic footprint, together with} the number of potential targets where \new{such TS} can be searched for, and whether or not \new{it is} continuous in time.

\end{abstract}


\end{frontmatter}



\section{Introduction} 
\label{sec:intro}

A variety of new developments in the last decade, such as the blossoming of exoplanet discovery and characterization science, as well as forthcoming improvements in observing capabilities from space and the ground, have opened new and exciting possibilities in the search for life in the Universe. The search for biosignatures, observational evidence of extraterrestrial life, has become one of the main science drivers for many new space missions and large telescopes. In parallel, these advances have motivated a renewed interest in the search for technosignatures (hereafter TS), defined as observational evidence for the existence of industry or technology in the Universe (for a more precies definition, see \cite{wright2018recommendations}).

In 2018, NASA began to consider including TS research as part of its research portfolio. In preparation for this, the agency organized a meeting (``NASA technosignatures workshop,''\footnote{https://www.hou.usra.edu/meetings/technosignatures2018/} held in September 2018 at the Lunar and Planetary Institute, Houston, USA) to learn about the current state of the art in the field. The workshop participants produced a report with an overall view of current possibilities in non-radio TS search \citep{technosignatures2018}. In August 2020, a second workshop was sponsored by NASA (``TechnoClimes 2020,'' Blue Marble Space Institute of Science\footnote{https://technoclimes.org/}) with the goal of producing a research agenda for TS science. This paper presents some of the conclusions of that workshop. Our aim is to establish an agenda of projects or space missions that would realize new avenues to search for TS. 

TS research has deep synergies with other areas of astronomy. It is often the case that TS science can be done in "commensal" mode, in which it takes advantage of data that are being acquired for other purposes (this is the same mode that had earlier received the less kind denomination of "parasitic", e.g. \cite{BZT83}). For instance, missions to observe exoplanets via transit photometry are also useful to search for transit TS at almost no extra cost. It is important to exploit such opportunities to the fullest, and another report from the TechnoClimes 2020 workshop explores such possibilities \citep{group2}. Here we take a different approach. We  present a set of projects and mission concepts that would explore new opportunities in this field, independent of any current limitations on funding or resources. The purpose of this approach is to provide an overview of the capabilities of current and near-term technology to conduct a systematic search for TS and present some ideas for conducting TS-specific searches. In the end, all of the concepts discussed in this paper would provide ancillary benefits in the form of extremely valuable data for other fields of astronomy, astrophysics and planetary sciences.

Even though we do not take them into consideration here, it is obvious that real-world limitations (mainly funding) are likely to play a major role in any prioritization of these concepts and, in fact, some exploratory work has been conducted in this direction\cite{LL19}. Those factors would need to be carefully balanced against the ancillary benefits for other areas of science and, more importantly, some inevitably uncertain guesses about the chances of success. For the latter, it would be important to consider the relevant TS in the context of the "nine axes of merit" \cite{S20}, as well as any other useful parameterization.

There are many different types of TS that might be produced by extraterrestrial technology and leave some observable evidence. Some may be intentional, such as radio or laser signals, transit beacons, or other intentional attempts at interstellar communication. Others are the byproduct of some industrial or engineering activity. Broadly speaking, for us to detect such signals at interstellar distances with our current sensitivities, such signals would need to be stronger than those produced by current human civilization, particularly the unintentional ones. It seems reasonable to assume that, as observing capabilities improve, TS searches would be increasingly sensitive to more moderate levels of technological activity. For instance, scientists in the 1960s were thinking about detecting star-system scale megastructures \citep{dyson1960search}, whereas modern technology could conceivably detect planetary-scale engineering \citep{gaidos2017transit}. We may view this as a ``technology balance" between the observer and the target technological capabilities. We are currently at the cusp of being able to detect our own TS at typical interstellar distances \citep{kopparapu2021,socas-navarro2018}. Only those species that have constructed or developed technologies much larger or more luminous than any of our own can be detected with our current astronomical infrastructure.  As this infrastructure becomes more sensitive, we will be able to detect technosignatures closer in scale to our own. It seems unlikely that civilizations with a relatively low level of technological development would enter into contact with each other, since that would require either very high sensitivities or highly visible engineering \cite{SNTS18} \cite{Barea2020}. This technology balance is the result of an observational bias by which "less advanced" civilizations lack the sensitivity needed to detect other civilizations unless they have built very large or luminous structures. In addition to this, a recent paper\cite{KFS20} presents an independent calcuation that leads to the conclusion that our first contact is likely to be with a more advanced civilization. The argument for this "contact inequality", as they call it, is based on statistical considerations about the lifespan of technological civilizations and does not take into account the observational bias. Therefore, these two independent arguments go in the same direction, reinforcing the notion that this is the most likely scenario for a first contact.

Perhaps an even more provocative idea is to consider a type of TS in the form of artifacts, for instance interstellar probes that might have been sent into the solar system a long time ago (potentially up to billions of years in the past), perhaps during a close encounter of our Sun with other stars. Such artifacts might have been captured by solar system bodies into stable orbits or they might even have crashed on planets, asteroids or moons \citep{a96,freitas1983search,benford2019looking}. Bodies with old surfaces such as those of the Moon or Mars might still exhibit evidence for such collisions. Systematic searches, which have not been conducted up to now, would at the very least provide upper limits on the existence of solar system artifacts  \citep{haqq2012likelihood,davies2013searching} and could do so at relatively low cost. 

Even negative results from volume-limited surveys for certain TS would be valuable because their absence may establish quantitative upper bounds on certain types of technologies or developmental stages of civilizations in the solar neighborhood. Such results may have multiple explanations including, but not limited to, the relative rarity of intelligence in the universe and may have profound implications for humanity's future \citep{2020GreatFilter}.

Nevertheless, such systematic observations would provide enormous ancillary benefits on solar system research and advance our knowledge about the objects being scrutinized. In this paper we consider a number of different mission concepts covering various types of TS. In addition to the TS research, we discuss other science goals that may be pursued with each one of these mission concepts.

\section{The $\iota$ vs $N_T$ diagram} 

The proposals put forward in this paper are very diverse in scope, size and targets. Lacking any better criteria, they are presented sorted by range. The variety of concepts, goals, targets and ancillary science,  makes it difficult to draw meaningful comparisons among them.

\begin{figure*}
    \centering
    \includegraphics[width=0.8\textwidth]{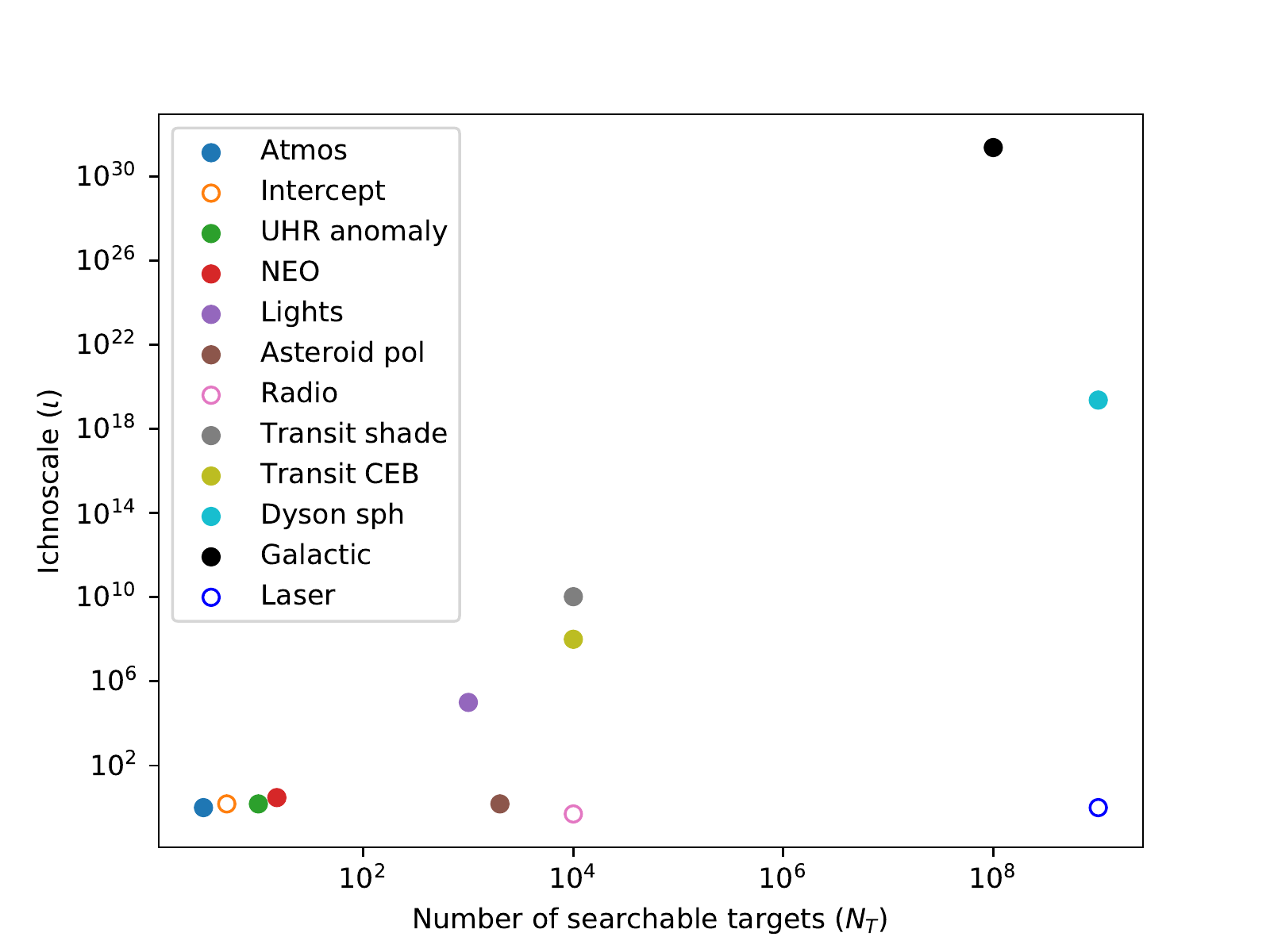}
    \caption{Ichnoscale (relative footprint of a given TS in units of current Earth technology) vs number of targets for several possible TS. Filled (empty) circles represent continuous (discontinuous) observables. See Table~\ref{tab:list} for label definitions.}
    \label{fig:ichnos}
\end{figure*}

We propose here a useful representation of TS that may be of general interest for future works in the field. Figure~\ref{fig:ichnos} shows an application to the TS discussed in this paper. In the $x$-axis we have the number of potential targets ($N_T$) where we can search for a given TS in a reasonable amount of time. For instance, this could be the number of stars we can observe in a large survey for radio signals, the number of exoplanets whose light curves we can analyze, etc. A search for artifacts on the Moon would have only one potential target, whereas a search for Dyson spheres in our galaxy would have of the order of a billion potential targets. Without any {\em a priori} knowledge or assumptions, TS with a larger $N_T$ (the right-hand side of the plot) are to be preferred. However, there are other important criteria, discussed below.

The $y$-axis of the plot has to do with the scale of the TS. Generally speaking, the detectability of a given TS is more difficult at larger distances. It is often argued that advanced civilizations are likely to build large geoengineering or stellar system engineering megastructures. This is not necessarily a two-way implication. There might be advanced civilizations that do not leave a large imprint on their planetary or space environment and will thus be very difficult to detect. And conversely, there might be younger civilizations that announce themselves very conspicuously, e.g. by broadcasting radio signals or releasing large amounts of industrial pollutants into their atmospheres. In fact, the term "advanced" is not a well defined one. What is really relevant in the context of TS searches is not how advanced a civilization is (whatever that could mean) but how large is its cosmic footprint, which is directly related to the detectability of its TS from a long distance. 

Instead of talking about more or less advanced civilizations, we introduce a parameter to quantify a TS footprint that we call ichnoscale\footnote{From the ancient greek root ichnos (\textgreek{ἴχνος}), which means footprint.} ($\iota$). This parameter is defined as the relative size scale of a given TS in units of the same TS produced by current Earth technology (hereafter CET). For instance, an alien orbital megastructure would have a $\iota$ parameter equal to the size of said megastructure divided by our largest object in orbit, which at present is the International Space Station. If the TS is a transit of such megastructure, then $\iota$ would be the ratio of their cross-sections. An alien probe sent into our solar system from a nearby star would have an ichnoscale slightly higher than one, since we are actually very close to having that capability, as evidenced by the development of the Breakthrough Starshot project\citep{parkin2018breakthrough}. In principle, we might expect that civilizations with a lower ichnoscale would be more abundant. Therefore, the best TS would be those appearing in the lower part of the plot in Figure~\ref{fig:ichnos}.

Finally, some TS are {\em continuous}, meaning that they will be observable all the time for the duration of our project. Once detected, they would be easily reproducible in observations by other groups with different instruments. For instance, a Dyson sphere would be continuous. Other TS are {\em discontinuous} and only exist for a relatively short period of time compared to the duration of a typical survey project. Examples would be transmissions. Another example could be a probe sent into our solar system to do a flyby. Detection of discontinuous TS requires some degree of serendipity and they might not be reproducible, for instance once the transmission has finished. Everything else being equal, continuous TS are more appealing than discontinuous ones. They are distinguished in the plot by using filled or empty circles. 

With all the above considerations, the ideal TS would be a filled circle at the lower-right corner of Figure~\ref{fig:ichnos}. Unfortunately, none of the points fulfill all three requirements simultaneously. We can envision TS that are either on the lower part of the figure, on the right-hand side, or some combination of both. Only the all-sky search for laser pulses is at the bottom right but it is a discontinuous TS.

\begin{table*}[]
    \centering
    \begin{tabular}{c|c|c|c}
        Technosignature &  Distance & Synergies & Label \\
         \hline
         Industrial gases on  & 10~pc  & Search for biosignatures. & Atmos  \\
         atmospheric spectra &  &  &  \\
         \hline
         Dark side  & 10~pc & Exoplanet surface  & Lights \\
         illumination &  &  characterization. &  \\
         \hline
         Starshades  & 1~kpc & Exoplanet characterization. & Transit shade \\
         in transit &    & Exomoons, exoring discovery. &  \\
         \hline
         Clarke exobelt & 1~kpc & Exoplanet characterization. & Transit CEB \\
         in transit &       & Exomoon, exoring discovery. &  \\
         \hline
         Laser pulses  & 100~kpc & Discovery of transients. & Laser \\
         \hline
         Heat from  & 1~kpc & IR astronomy. & Dyson sph \\
         Dyson spheres  &  &   &  \\
         \hline
         Heat from galactic   & Mpc & IR astronomy. & Galactic \\
          civilization  &  &   &  \\
         \hline    
         Radio signals  & 10~kpc & Radioastronomy. & Radio \\
         \hline
         Artifacts on   & 1~AU & NEO exploration.  & NEO \\
         Earth co-orbitals  &   & Planetary defense. &  \\
         \hline
         Artifacts on Moon    & 1~AU & Planetary geology. & UHR anomaly \\
         and other surfaces  &     &  &  \\   
         \hline
         Interstellar  & 1~AU & Solar system. & Intercept \\
         probes             &  & Interstellar objects. & \\
         \hline
         Artifacts on  & 1~AU & Asteroids. Comets. Planetary & Asteroid pol \\
          asteroids &  & defense. Interstellar objects. &   \\
         \hline
    \end{tabular}
    \caption{List of TS discussed with a detection distance scale, a summary of possible ancillary benefits of a mission to explore them and the corresponding label in Fig~\ref{fig:ichnos}}
    \label{tab:list}
\end{table*}

The relation between a TS ichnoscale and the age of its civilization is not necessarily linear. In fact, experience with CET suggests that it may grow exponentially in some cases. For instance \cite{socas-navarro2018} suggests that the $\iota$ of our Clarke belt has been growing exponentially over the past 20 years. If this exponential growth were to be sustained for a much longer period of time, $\iota=10^8$ would be reached in about 200 years. We do not know if such rapid growth will be sustained, of course, but in principle it seems well within our capabilities. Whether it will happen or not is probably a matter of motivation rather than feasibility.

All of the proposals discussed in this paper lie roughly below or very close to the diagonal of the representation in Fig~\ref{fig:ichnos}, which is the "good domain" of the plot. This is the region where TS have a combination of high number of search targets or a relatively low ichnoscale. Both factors would be expected to increase the chances of success, assuming that low-$\iota$ civilizations are more abundant. Above the diagonal is the "bad domain", where TS require a high-$\iota$ (presumably less likely) and we would have few places to search. That region is empty in our figure. The distribution of our proposals in the $\iota$ vs $N_T$ diagram shows that none of the proposed searches is clearly superior to the others. They all offer different trade-offs in terms of advantages and drawbacks, which are discussed in the following sections.

\section{New searches on existing data}

TS science could greatly benefit from support for searches of data acquired with existing instrumentation or expected from upcoming facilities. Many TS might potentially be found in exoplanet observations, both in transit photometry and spectroscopy. Upcoming missions with coronographic and/or direct exoplanet observing capabilities will provide unprecedented data that could be mined for possible signs of technology.

Forward modeling to simulate TS observations would produce a useful catalog of possible light curves, atmospheric spectral features and other observables to check against the data. This includes, but is not limited to, the following:
\begin{itemize}

\item Complex light curve analysis, including objects in orbit around exoplanets \citep[e.g.,][]{gaidos2017transit}. Exomoon and exoring discovery and characterization are likely ancillary benefits from this work.

\item Atmospheric spectral signatures in transmission and reflection on exoplanet atmospheres and surfaces \citep{slf10,lingam2017natural}. The data analysis is likely useful for projects searching for biosignatures but the simulation and modeling of relevant TS is an independent effort.

\item Photometric and spectroscopic searches for night-time illumination \citep{slf10,loeb2012detection}. These projects would have synergies with exoplanet characterization and surface mapping.

\item Network analysis methods are already being explored in the characterization of biosignatures \citep{walker2018exoplanet} and could help to narrow the search for TS. One example under development is a network fitting algorithm to determine an optimal navigation-communication pathway among a sample of exoplanets in a given region of interstellar space. A region of space with the right distribution of suitable worlds to become a communication hub may be a promising place to search. TS might be more abundant there, just like Earth TS are more abundant wherever there is a high density of human population, which in turn tends to clutter in the form network structures\cite{Davis19,davis_etal2020}.


\end{itemize}

\section{Observing the Spectra of Planetary Atmospheres}

The history of life on Earth provides a starting point in the search for biosignatures on exoplanets \citep{krissansen2018disequilibrium,Pall2018,fu03200x},  with the various stages of Earth’s evolution through the Hadean (4.6 -4 Gyr), Archean (4 - 2.5 Gyr), Proterozoic (2.5 - 0.54 Gyr), and Phanerozoic (0.54 Gyr - present) eons representing atmospheric compositions to use as examples of spectral signatures of an inhabited planet.  By extension, the search for technosignatures likewise can consider Earth’s evolution into the Anthropocene epoch \citep{crutzen2006anthropocene,lewis2015defining,frank2017earth} as  a  template  for  future  observing  campaigns that  seek  to  detect  evidence  of  extraterrestrial  technology.   For instance, \citet{lin2014detecting} discussed the possibility of detecting tetrafluoromethane (CF$_4$) and trichlorofluoromethane (CCl$_3$F) signatures in the atmospheres of transiting Earths around white dwarfs with JWST, which the authors calculate could be detectable if these compounds are present at 10 times the present Earth level.  Similarly, \citet{kopparapu2021} showed that present Earth-level of nitrogen dioxide (NO$_{2}$), produced as  a  byproduct  of  combustion or nuclear technology, could be detectable on an Earth-like planet around a Sun-like star at 10\,pc with $\sim$400 hours of observation time with a telescope like the Large Ultraviolet Optical Infrared Surveyor (LUVOIR, \citet{Fischer2019}). Some other projects for future missions to observe in the mid-infrared range might also enable this type of observations, such as the proposed Habitable Exoplanet (HabEx, \citet{Gaudi2019}), Origins Space Telescope (OST, \citet{Cooray2019}), or Large Interferometer for Exoplanets (LIFE, \citet{Quanz2020life}) missions.


Atmospheric spectral searches for industrial technology would be ``passive'' in the sense that the civilizations do not need to actively do anything to communicate with us. Furthermore, at least for remote exoplanet observations, the near/medium term missions include instruments that perform atmospheric spectral characterization, which is ideal for piggybacking technosignature science along with routine observations. Despite these advantages, there has been no major work performed to explore atmospheric technosignatures, other than the two studies mentioned above. Nevertheless, we can speculate about what kind of instruments and wavelength coverage we need to specifically observe the above mentioned industrial gases in the atmospheres of exoplanets.

CF$_4$ and CCl$_3$F have absorption bands in the mid-IR region between $7.76 - 7.84 \,\mu$m and $11.6 - 12 \,\mu$m, respectively. CF$_4$ is a very narrow spectral line, with a width of $\sim 0.08 \mu$m. Therefore a high-resolution instrument with R $>100$ in the mid-IR can resolve the feature \citep{schneider2010far}. The MIRI instrument on {\it JWST} ($5 - 28 \mu$m) already provides medium resolution spectroscopy (MRS) with R from 1500 to 3500. Similarly, CCl$_3$F has a spectral line width of $\sim 0.4\mu$m requiring a modest R $> 30$. While the resolution seems already achievable, the concentration and the exposure times require  large values. \citet{lin2014detecting} show that to detect these CFC gases on an Earth-like planet around a white-dwarf star with JWST in one day with a signal-to-noise ratio (SNR) of 5, the concentrations need to be 10 times the present Earth-level. Furthermore, CCl$_3$F (CFC-11) and CCl$_2$F$_2$ (CFC-12) are another set of atmospheric CFC pollutants that absorb between 8-14 micron mid-IR wavelengths. These are two of the most abundant CFCs in Earth's atmosphere with elevated levels that have persisted despite the Montreal Protocol\cite{MDY+18,WWS00}.

There has not been a follow-up study that explored further on the detectability of these features on an Earth-like planet around a M-dwarf star with either {\it JWST} or the flagship mission concept Origins Space Telescope {\it OST}. Such studies are sorely needed so that instrument capabilities can be developed.

NO$_{2}$ has broad absorption  between  $0.25-0.6 \,\mu$m,  which  has  little  overlap  with absorption from other terrestrial atmospheric constituents.  Absorption features are also present at 3.5$\,\mu$m, 6.4$\,\mu$m and 10-16$\,\mu$m, but  these  overlap 
with  absorption  bands  from  H$_{2}$O,  CO$_{2}$. Simulations by \citet{kopparapu2021} indicate that direct imaging spectroscopic observations with LUVOIR-15m telescope in the near UV (NUV, 0.2–0.525$\,\mu$m) and visible (VIS, 0.515–1.030$\,\mu$m ) channels can detect current Earth-level NO$_{2}$ absorption features, if present, in $\sim 1000$ hours of observation time with a SNR of 5 at 10pc distance. For comparison, Hubble Space Telescope's large programs such as the Ultra Deep Field (UDF) and CANDLES surveys used between- 400-900 hours of observation time over a period of 1-3 years. The resolution needed in the NUV can be very low as the NUV band is quite broad. \citet{kopparapu2021} used R = 6 for NUV and R = 70 for the VIS region to maximize the SNR. They also argue that Earth-like planets around K-dwarfs may have a slight advantage in detecting NO$_{2}$ because of comparatively fewer shorter wavelength photons that destroy NO$_{2}$. 

The above discussion indicates that the instrument technology to search for atmospheric technosignatures either already exists, or is within reach. We are at the cusp of detecting spectral fingerprints of extraterrestrial technology. What this particular sub-topic needs is dedicated time to collect the data for this high-risk, high-reward endeavor. A nice advantage of this method of detecting atmospheric technosignatures is that the same instruments and telescopes can be used to characterize atmospheres of exoplanets. Our view of habitability and technosignatures is based on our own Earth's evolutionary history. There are innumerable examples in the history of science where new phenomena were discovered serendipitously. By having a dedicated mission to look for atmospheric technosignatures that also covers exoplanet science, we can increase our chances of detecting extraterrestrial technology on an unexpected exoplanet, or may discover a spectral signature that we usually do not associate with technology. The only way to know is to search.

\section{All Sky Laser Searches}

Pulsed lasers are a promising TS, proposed by \citet{schwartz1961interstellar} shortly after the development of the first laser.  After radio searches, laser searches are the best developed TS search strategy, thanks to the work of Paul Horowitz over many decades. 

Today, the premier laser search strategy is employed by PANOSETI \citep{PANOSETI_SPIE,PANOSETI_AAS} Because the signals sought are pulsed, they are broadband, obviating the need to choose a single frequency range, so a search may proceed over the entire optical band or wide regions of the infrared with a single instrument. Because there are no natural sources of short pulses (or even any artificial ones at sufficiently short pulse durations) there is no background sources and no confounders to contend with. PANOSETI is able to search most of the available sky above its horizon all night long simultaneously and with good coincidence detection to rule out instrumental false positives. An intense laser pulse within the capabilities of CET could be detectable across the galaxy.

Extending this search into infrared wavelengths would improve the search space by an order of magnitude, but requires fast infrared detectors capable of nanosecond timing.  Such detectors currently exist but are generally very expensive and consist of a small number of pixels, insufficient for an all-sky search. A dedicated technology development effort to improve these detectors would enable a space mission to search the entire fast infrared sky, and advance time domain astronomy generally by being sensitive to bright transients at all timescales, including potentially pulsars, fast radio bursts, and new phenomena \citep[e.g.,][]{gajjar2018highest,lacki2020one}.

\section{Waste heat mission}

Studies of technological waste heat (e.g. searches for Dyson spheres) benefit most from all-sky surveys at mid- and far-infrared wavelengths \citep{2020SerAJ.200....1W}.  The launch of IRAS, sensitive out to 100 microns, led to some optimism that Dyson Spheres might be detected, but the discovery of the infrared cirrus revealed that IRAS had background-limited source sensitivity (not photon-limited). Because its angular resolution was quite poor, this resulted in a much lower sensitivity to Dyson spheres than naive estimates suggested.

Nonetheless \cite{TKP00} and \cite{carrigan2009iras} analyzed the data of the brightest sources using the IRAS spectrometer and put the first upper strong limits on Galactic Dyson spheres.  Since then, the WISE mission has provided an all-sky survey out to 24 microns, which motivated a further search for Dysonian technosignatures \citep{wright2014g,wright2014g-2,griffith2015g,wright2016g}.  Unfortunately, without sensitive longer-wavelength data of the hundreds of millions of sources detected by WISE it is difficult to construct SEDs of most sources and perform a rigorous search. The poor angular resolution of IRAS also means it can be challenging to confidently match IRAS sources to optical and NIR sources.

An update to the IRAS (and Akari) missions with improved point source sensitivity would help tremendously, and provide major scientific return beyond TS search.  This would require two primary improvements over IRAS: a larger primary mirror (for better angular resolution) and superior IR detectors (for better calibration and sensitivity).  The first of these is probably the most challenging: at 100 microns, one needs a 2.5\,m mirror to have better than 10” resolution.  For scale, the ESA Herschel mission was a mid- to far-infrared 3.5\,m space telescope (that performed targeted observations with a suite of instruments) and cost approximately a billion USD.

A prior concept study would be desirable to determine the feasibility, cost, and broader scientific impact of a successor to IRAS: a 3\,m class space telescope performing an all-sky survey with very good point source sensitivity at many infrared bands out to at least 100 microns, similar to IRAS. The benefits to core astrophysics for such a mission might be numerous (studies of dust, exozodi, brown dwarfs, protoplanetary disks, etc) and they would be synergistic with the capabilities of future infrared flagships such as Origins\cite{L20}, which could target specific objects of interests identified by such a survey. Such a mission would be sensitive to Dyson spheres at up to kpc distances.

\section{Radio observatory on the far side of the Moon}

The scientific search for TS began with the suggestion by \citet{cocconi1959searching} to search for radio transmissions originating from extraterrestrial technology. The first search for narrow-band radio waves was conducted shortly afterwards\cite{drake1961project}, which led to the birth of radio SETI as a discipline. Radio searches for TS today include observations by the SETI Institute's Allen Telescope Array \citep{tarter2011first,harp2016seti}, the Breakthrough Listen survey that includes several radio and optical facilities \citep{isaacson2017breakthrough,lipman2019breakthrough,wlodarczyk2020extending}, and numerous other efforts by research groups around the world  \citep[e.g.,][]{montebugnoli2010seti,rampadarath2012first,zhang2020first}.

The far side of the Moon is of great interest because it is nearly free of contamination from human radio emissions. A permanent radio telescope facility in this region would permit radio searches of unprecedented sensitivity\citep{michaud2020lunar} . Earth radio waves with frequencies as low as 50~kHz would be attenuated by at least 10 orders of magnitude. Of course such a facility would be very relevant for other astronomical purposes as well and is well aligned with NASA's priorities in the continued human exploration and development of the Moon.

In particular, the band between 70-80 MHz is particularly critical for cosmology. This is where the cosmological redshift should have brought the 21cm line formed during the ``cosmic dawn.'' the time when the first stars were formed. Observing this frequency from Earth is very difficult because of intererence with radio stations. A first detection of the cosmic signal may have been obtained by the EDGES project \citep{edges2018} but it still remains controversial \citep[see e.g.,][]{Hills2018}.

In parallel with such a project, there should be a strong push by the community for a special radio protection of the far side of the Moon \citep{maccone2008protected,maccone2019moon}. Radio observations may be complicated by future artificial satellites that may be launched in support of space missions. An example is China's Queqiao satellite (in support of the Chang'e 4 mission), which has orbited the Earth-Moon L2 Lagrange point since May 2018 and represents the only current source of radio contamination on the lunar farside. Contamination by Queqiao is probably not too harmful by itself but the possibility of other national and commercial space agencies also placing satellites at Earth-Moon L2 heightens the need to protect this unique radio-quiet location for general radio astronomy (including TS searches).

\section{Exploration of Near-Earth Objects}

Stars do not remain in fixed positions. In addition to the bulk orbital motion around the galaxy, nearby stars appear to us as zooming past in various directions. The solar system undergoes relatively frequent (on cosmic scales) close stellar encounters, typically a star penetrates our Oort cloud (coming within a light-year from the Sun) every 10$^5$ years. This means that, since the beginning of life on Earth, there have been tens of thousands of such close encounters. An extraterrestrial civilization that passes nearby might notice that there is an ecosystem here, due to the out-of-equilibrium atmospheric chemistry.

If our civilization is already considering the launch of an interstellar probe (Breakthrough Starshot, \cite{parkin2018breakthrough}) to the nearest star at a comparable distance, it is not unreasonable to think that perhaps in one or more of those close encounters some other species might have sent probes to explore our solar system. This idea was first raised by ref\cite{bracewell1960communications} just one year after the suggestion to search for radio TS\cite{cocconi1959searching}; however, only a handful of searches for such objects have been conducted \citep{freitas1980search,valdes1983search}.

We suggest that near-Earth objects (NEOs) could be examined for evidence of artifacts, possibly even “lurkers” \citep{benford2019looking,bracewell1960communications}.  NEOs provide an ideal post to watch our world from a secure natural
object, providing resources an ETI might need: materials, a firm anchor, concealment. Yet their surfaces have been little studied by astronomy and not at all by SETI or planetary radar observations. The recently discovered group of nearby Earth co-orbital objects is an attractive location as well. 

Close inspection of bodies in these regions can now be done with XXI century observatories and spacecraft. The great virtue of searching for lurkers is their lingering endurance in space, long after they go dead. The Moon and the Earth Trojans have a greater probability of success than the co-orbitals \citep{benford2020a,benford2020b}. These minor bodies may hold clues on the formation and evolution of our solar system, which makes them interesting objects of study.

\section{Ultra High-Res imaging with on-board AI for anomaly detection}

Following the rationale of close encounters with other stars on time scales of 100,000 years, it would be desirable to explore the older surfaces in the inner solar system for possible evidence of technological artifacts, whether they were sent intentionally there or simply ended up colliding with a major body after their mission. The Moon and Mars  are attractive in this context as possible locations where such devices might have ended up and little surface evolution takes place. Evidence of impacts or existing artifacts might be preserved for between millions and billions of years depending on conditions. Other attractive bodies might be Mercury, Ceres or the largest asteroids.

A global high-res mapping of the Moon is constructed on a monthly basis by the Lunar Reconnaissance Orbiter (LRO) with a resolution of 100m/pixel. At this resolution some human artifacts such as Apollo landing sites may be recognized but smaller and/or less obvious TS would be missed. The probe has a higher resolution mode of 0.5m/pixel but only special locations have been mapped in this mode.

Ultra high-resolution (at the $\sim$10cm per pixel level) mapping of the entire surface would be desirable in the search for TS. Such UHR is currently technologically feasible but sending back such enormous amounts of data would require a colossal telemetry capacity. However, new and exciting possibilities are now available with machine learning techniques, which have been demonstrated to be useful in processing similar datasets to identify anomalies. In particular, \citet{lesnikowski2020unsupervised} have successfully trained an artificial neural network to search agnostically for anomalies in the LRO maps. 

We propose a probe to explore the lunar surface in UHR. Assuming an altitude similar to LRO, a resolution of $\sim$10cm/pixel would correspond to about 0.4 arc-seconds and could be achieved with a telescope of ~40cm aperture in the visible. At this resolution, spacecraft motion limits the maximum exposure time to the order of 10$^{-4}$ seconds. The expected signal-to-noise ratio achievable by such a system in broad-band images would be approximately 300.

Instead of storing and sending the data back to Earth, an on-board artificial intelligence would conduct a real-time analysis to select the most interesting ones (highest anomaly ranking). The images selected would be sent back to Earth for further analysis along with context imaging provided by lower resolution, wider field cameras. A buffer of UHR images would be maintained so that a strip in the scanning direction can be sent prior and after the high-anomaly images. Additionally, further passes over the same region would be used to construct a UHR mosaic. 

A sun-synchronous orbit would be desirable to scan the surface with a consistent illumination. Since the Moon orbits around Earth, it is not possible to maintain a strict sun synchronicity but its variation will be slow because the probe orbital period is much shorter than the Moon’s orbit around Earth. This implies that illumination will be consistent between consecutive passes, allowing for the construction of mosaics, but will vary slowly on time scales of weeks. It will then be possible to observe the same region with a different illumination at later times. A similar mission to Mars might be considered. The time scales for martian dust to cover different size structures would need to be evaluated.

These UHR missions with on-board anomaly detection would be extremely useful not only in TS search but also for planetary geology. It would open the opportunity to observe planetary surfaces in unprecedented detail and to identify what may be the most interesting sites for study, boosting the chances for new discoveries or identifying optimal landing sites. New technological developments would be needed for the implementation of the required on-board real-time processing capabilities. Ideally, one would like to examine $\sim$10 images per second in order to cover the entire field as it is observed. Alternatively, there could be some degree of frame dropping and the scanning would progress along multiple passages over the same area. All these developments would undoubtedly open new possibilities for future space missions.

\section{Ready to launch intercept mission}

One of the most important developments in astrophysical research for the upcoming decades is likely to be the time domain astronomy enabled by all-sky synoptic surveys such as the Vera C. Rubin Observatory. It is expected that this facility will identify several interstellar interlopers per year, objects possibly similar to 1I/‘Oumuamua for 2I/Borisov. It is very difficult to observe such objects in detail because of the short lead time since discovery to optimal observing conditions. For instance,1I/‘Oumuamua was detected as it was already on its way out of the solar system. 

There has been some speculation about the possible artificial nature of ‘Oumuamua \citep{Bialy_2018}, given its aspect ratio, anomalous acceleration and absence of observed outgassing. While this proposal has been challenged by several authors \citep{oumuamua_ISSI_2019}, the fact is that we lack the necessary data to conclusively establish its nature and origins. It is frustrating to think that such a (possibly) unique object passed so close without a chance for close scrutiny. 

We propose an intercept mission for interesting interstellar interlopers that would be ready to launch in case a target of opportunity presents itself. The mission could be a conventional chemical-propulsion spacecraft equipped with general purpose photometric and spectroscopic instrumentation. If the target is detected with sufficient lead time, thanks to the new survey facilities, it may be possible to catch it within 20 years \citep{hibberd2020project}.

Alternatively, a more ambitious mission using light sail technologies could send a sailcraft to catch the interstellar object in a much shorter time. This mission concept could be conceived in collaboration with Breakthrough Starshot \citep{parkin2018breakthrough} as a smaller pathfinder project.

\section{Asteroid polarimetry mission}

As mentioned above, roughly every 100,000 years a star comes within nearly a light-year from the Sun, providing tens of thousands of opportunities for technologies similar to ours to \new{have launched} probes into our solar system. 

Artificial objects made by human technology usually have high reflectivity and imprint strong linear polarization on reflected light. This is mostly a consequence of construction with very flat metallic surfaces. Natural objects in our space environment, such as comets or asteroids, rarely produce any measurable polarization. We see merit in a new mission to explore the minor bodies of the solar system with emphasis on polarimetry, on the rationale that artificial objects may stand out from the background in polarization. The main target would be the study of the smaller objects, of sizes below 10\,m, which cannot be observed from Earth. As ancillary benefits, we would obtain a robust statistical description of the smaller objects in the solar system, of which very little is currently known. 

There are two possible approaches that we recommend. The first is a mission into the asteroid belt. The second one aims for the Jupiter trojans. The gravitational well of Jupiter might act as a fishing net to capture objects, both natural and artificial, coming from outside the solar system. We recommend a preparatory theoretical study to explore the dynamics of objects coming from beyond the solar system to determine the most interesting possible graveyards where such objects might end up if captured. Interstellar objects are of great interest for several reasons and it is likely that many of them have been captured and are can be found abundantly in the solar system\cite{HD20}. The study of these objects has deep implications in astrophysics and astrobiology\cite{LL18}.

Asteroid Belt mission: A telescope similar to Kepler would be sensitive to objects of 10\,m up to a distance of 0.02\,AU (assuming a high albedo of ~0.8) or ~0.01\,AU for typical asteroid albedos. Extrapolating the current knowledge of asteroid size distribution, there should be some 250,000 asteroids of ~10\,m in the radius of 0.02\,AU accessible to such telescope in the asteroid belt. The mission could be designed with an elliptical orbit having the perihelion near the Earth’s orbit and the aphelion in the asteroid belt. Under these conditions it would regularly dive into a different region of the belt, probing a different space in every orbit.

Jupiter Trojans mission: Given that most of the fuel and the cost of a mission launch is spent during the first phase, it would be a relatively economic concept to launch two identical probes that would be separated during a gravity assist maneuver to reach the two trojan spots of Jupiter. The main drawback is that, being nearly a factor 2 further away from the Sun, the visibility of any possible objects would be reduced.

The information gathered on the abundance and properties of smaller bodies would be of great relevance for Earth protection against asteroid impacts.

\section{Conclusions}

The field of TS, albeit young, already has a rich history of innovative ideas to search for potential signs of extraterrestrial technology. The idea of Dyson spheres dates back to the 1960s. In the 1980s, Harris was thinking about remnants of interstellar propulsion and estimated the detectability of their gamma-ray emissions\cite{H86} . The search for TS deals with questions that have profound implications on the future of humanity. Perhaps one the most important is whether technological civilizations are ephemeral or, on the contrary, can be long lasting. A closely related question is whether space faring civilizations are common, and if humankind will eventually become \new{one of them. We do not yet} have any answers for these and other important questions but if we can start to explore the search parameter space, even in the absence of any detection we may be able to gain some valuable insights.

Given the recent technological and scientific developments in astrophysics, exoplanet science and astrobiology, \new{it seems to us} that the present time offers a perfect opportunity for a revitalization of the research on TS.
The TechnoClimes2020 workshop continued the work started in the Technosignatures 2018 meeting to bring together a community of scientists with diverse backgrounds and discuss their vision for the future of the field. These workshops have highlighted a consensus in the idea that new opportunities are opening to search for TS. This paper reports on the reflections of the community, as represented by the participants, on what new avenues might be pursued to optimize the chances of success.

We have introduced a novel framework that might also be of interest for future work dealing with multiple classes of TS of widely different scales and nature. This framework operates by \new{defining} the ichnoscale ($\iota$) \new{and analyzing its relation to} the number of targets ($N_T$) and the persistence of the signal. The ichnoscale is a quantification of how conspicuous a given TS is, compared to current Earth technology. It is a useful parameter to consider in SETI discussions. An important advantage is that it frees us from having to use vague and ill-defined terms such as "advanced civilization". A $\iota$ vs $N_T$ diagram, such as the one presented in Fig~\ref{fig:ichnos}, might be a valuable tool for the study of TS.

The mission concepts sketched in the various sections of this paper would explore new regions of the parameter space that have not been probed in the past.
Even if no TS are found, the observations acquired would be useful in many other areas, ranging from solar system exploration to radioastronomy or cosmology. Since much of the TS science is driven by search of anomalies, these missions are expected to present a high potential for discovery of new phenomena. 

\section*{Acknowledgements}

This study resulted from the TechnoClimes workshop (August 3-7, 2020, technoclimes.org), which was supported by the NASA Exobiology program under award 80NSSC20K1109. The authors are grateful to Joseph Lazio for comments on an earlier version of the manuscript and Maria Ribes Lafoz for expert linguistic advise that led to the term "ichnoscale". HSN acknowledges support from the Spanish Ministerio de Ciencia, Innovaci\'on y Universidades through project PGC2018-102108-B-I00 and FEDER funds. The Center for Exoplanets and Habitable Worlds and the Penn State Extraterrestrial Intelligence Center are supported by the Pennsylvania State University, the Eberly College of Science, and the Pennsylvania Space Grant Consortium. Any opinions, findings, and conclusions or recommendations expressed in this material are those of the authors and do not necessarily reflect the views of their employers, NASA, or any other sponsoring organization.

\section*{Data Statement}
No new data were generated or analysed in support of this research.


\bibliography{main}

\end{document}